# Fastest Distributed Consensus on Path Network

Saber Jafarizadeh
Department of Electrical Engineering
Sharif University of Technology, Azadi Ave, Tehran, Iran
Email: jafarizadeh@ee.sharif.edu

*Abstract*— **Providing an analytical solution for the problem of finding Fastest Distributed Consensus (FDC) is one of the challenging problems in the field of sensor networks. Most of the methods proposed so far deal with the FDC averaging algorithm problem by numerical convex optimization methods and in general no closed-form solution for finding FDC has been offered up to now except in [3] where the conjectured answer for path has been proved. Here in this work we present an analytical solution for the problem of Fastest Distributed Consensus for the Path network using semidefinite programming particularly solving the slackness conditions, where the optimal weights are obtained by inductive comparing of the characteristic polynomials initiated by slackness conditions.**

## I. INTRODUCTION

A sensor network consists of a large number of sensors, called nodes, which are densely deployed inside or close to a phenomenon. The main goal of using sensor networks is to reach a global decision or estimation about the state of the phenomenon reducing the average probability of error. These nodes are capable to communicate between each other in a cooperative manner and over relatively short distances, a limitation that has to be considered is that making global decisions (i.e., based on information collected by all sensors) has to be done through local communication between neighboring sensors, this problem is widely known as distributed detection in sensor networks. One of commonly used methods in this issue is distributed consensus averaging algorithm [1, 2] which is gaining considerable attention nowadays and one of the main research directions in this field is the computation of the optimal weights that yield the fastest convergence rate to the asymptotic solution [1] which is known as Fastest Distributed Consensus (FDC). Providing an analytical solution for the problem of finding Fastest Distributed Consensus is one of the challenging problems in the field of sensor networks. Most of the methods proposed so far deal with the Fastest Distributed Consensus problem by numerical convex optimization methods and in general no closed-form solution for finding Fastest Distributed Consensus has been offered so far except in [3, 4] where the conjectured answer by [5] for Path network has been proved. In [6], the author proposes an analytical solution for the problem of finding Fastest Distributed Consensus based on Semidefinite Programming (SDP), by imposing the slackness conditions and comparing the corresponding characteristic polynomials of blocks of weight matrix for two networks, which are containing Path as a particular case.

Here in this work we present an analytical solution for the problem of Fastest Distributed Consensus for the Path network using semidefinite programming particularly solving the slackness conditions, where the optimal weights are obtained by inductive comparing of the characteristic polynomials initiated by slackness conditions.

For notational simplification and due to lack of space, we illustrate our approach with a specific graph. However, in [6] author has shown that this method can be applied to networks with more general topologies.

The organization of the paper is as follows. In section II we present the network and state the problem. Section III is containing the proposed methods and main results of the paper, namely the exact determination of optimal weights for fastest distribution consensus algorithm via semidefinite programming in Path networks and finally section IV presents the conclusion and topics for future research.

## II. PROBLEM STATEMENT

We consider a path network with $n \geq 2$ nodes, labeled $i = 1, \dots, n$ and $n - 1$ edges connecting pairs of adjacent nodes, as shown in Fig. 1

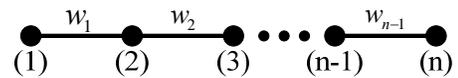

Fig.1. Weighted Path network with $n$ vertices

Each node $i$ holds an initial scalar value $x_i(0) \in \mathcal{R}$, and $x(0) = (x_1(0), \dots, x_n(0))$ denotes the vector of initial node values on the network. The main purpose of distributed consensus averaging is to compute the average of the initial values, $(1/n)\sum_{i=1}^{n} x_i(0)$ via a distributed algorithm, in which the nodes only communicate with their neighbors. In this work, we consider distributed linear iterations, which have the form

$$x(t+1) = Wx(t) \qquad (2\text{-}1)$$

where $t = 0,1,2,\dots$ is the discrete time index and $W$ is the network's weight matrix defined as

$$W = \begin{bmatrix} 1-w_1 & w_1 & \cdots & 0 \\ w_1 & 1-w_1-w_2 & \cdots & \vdots \\ \vdots & \vdots & \ddots & w_{n-1} \\ 0 & \cdots & w_{n-1} & 1-w_{n-1} \end{bmatrix} \quad (2\text{-}2)$$

which is a symmetric tridiagonal matrix due to the topology of network's undirected associated graph and also $W$ has the same sparsity pattern as the adjacency matrix of the network's associated graph. In [1] fastest distributed consensus problem has been formulated as the following minimization problem

$$\begin{aligned} \min_{W} \quad & \max(\lambda_2, -\lambda_n) \\ s.t. \quad & W = W^T, W\mathbf{1} = \mathbf{1} \\ & \forall (i,j) \notin E: W_{i,j} = 0 \end{aligned} \quad (2\text{-}3)$$

Where $\mathbf{1}$ is the column vector of all one and $1 = \lambda_1 \geq \lambda_2 \geq \cdots \geq \lambda_n \geq -1$ are eigenvalues of $W$ arranged in decreasing order and $\max(\lambda_2, -\lambda_n)$ is the *Second Largest Eigenvalue Modulus* (SLEM) of $W$.

### III. MATH

The main problem (2-3) can be formulated in the semidefinite programming form as [1]:

$$\begin{aligned} \min_{W} \quad & s \\ s.t. \quad & -sI \preccurlyeq W - \mathbf{1}\mathbf{1}^T/n \preccurlyeq sI \\ & W = W^T, W\mathbf{1} = \mathbf{1} \\ & \forall (i,j) \notin E: W_{i,j} = 0 \end{aligned} \quad (3\text{-}1)$$

Introducing linear independent bases

$$\alpha_{i,j} = \begin{cases} 1/\sqrt{2} & j = i \\ -1/\sqrt{2} & j = i+1 \\ 0 & \text{otherwise} \end{cases} \text{ for } \begin{aligned} i &= 1, \ldots, n-1 \\ j &= 1, \ldots, n \end{aligned} \quad (3\text{-}2)$$

the weight matrix can be written as

$$W = I - \sum_{i=1}^{n-1} 2w_i \alpha_i \alpha_i^T \quad (3\text{-}3)$$

Based on [1] and considering the weight matrix as in (2-2), one can express fastest distributed consensus problem for Path network in the form of semidefinite programming as:

$$\begin{aligned} \min_{W} \quad & s \\ s.t. \quad & -sI \preccurlyeq W - \mathbf{1}\mathbf{1}^T/n \preccurlyeq sI \end{aligned} \quad (3\text{-}4)$$

By substituting expansion (3-3) in the constraints of (3-4) we have

$$sI + \mathbf{1}\mathbf{1}^T/n - I + \sum_{i=1}^{n-1} 2w_i \alpha_i \alpha_i^T \geq 0 \quad (3\text{-}5)$$

$$sI - \mathbf{1}\mathbf{1}^T/n + I - \sum_{i=1}^{n-1} 2w_i \alpha_i \alpha_i^T \geq 0 \quad (3\text{-}6)$$

In [6] the standard primal semidefinite programming problem has been defined as

$$\begin{aligned} \min \quad & c^T x \\ s.t. \quad & F(x) \geq 0 \end{aligned} \quad (3\text{-}7)$$

where $c$ is a given vector and components of vector $x^T = (x_1, \ldots, x_n)$ are the variables of the problem, and $F(x) = F_0 + \sum_i x_i F_i$, for some fixed hermitian matrices $F_i$ and the inequality sign in $F(x) \geq 0$ means that $F(x)$ is positive semidefinite. In order to formulate problem (3-4) in the form of standard semidefinite programming (3-7), we define $F_i, c_i$ and $x$ as below:

$$\begin{aligned} F_0 &= -\sigma_z \otimes (I - \mathbf{1}\mathbf{1}^T/n) \\ F_i &= \sigma_z \otimes \alpha_i \alpha_i^T, \quad i = 1, \ldots, n-1 \\ F_n &= I_{2n} \\ c_i &= 0, \quad i = 1, \ldots, n-1 \\ c_n &= 1 \\ x^T &= [2w_1, 2w_2, \ldots, 2w_{n-1}, s] \end{aligned} \quad (3\text{-}8)$$

where $\sigma_z = \begin{bmatrix} 1 & 0 \\ 0 & -1 \end{bmatrix}$ and the associated dual program is defined as [6]:

$$\begin{aligned} \max \quad & -Tr[F_0 Z] \\ s.t. \quad & Z \geq 0 \\ & Tr[F_i Z] = c_i \end{aligned} \quad (3\text{-}9)$$

where the data $c$ and $F_i$ are the same as in (3-8) and the variable $Z$ is the real symmetric (or Hermitean) positive matrix. We define $Z$ as

$$Z = \begin{bmatrix} z_1 \\ z_2 \end{bmatrix} \cdot [z_1^T \quad z_2^T] \quad (3\text{-}10)$$

where $z_1$ and $z_2$ are n-array column vectors and obviously (3-10) choice of $Z$ implies that it is positive definite. Using the constraints $Tr[F_i Z] = c_i$ we obtain

$$(z_1^T \alpha_i)^2 = (z_2^T \alpha_i)^2, \quad i = 1, \ldots, n-1 \quad (3\text{-}11)$$

$$z_1^T z_1 + z_2^T z_2 = 1 \quad (3\text{-}12)$$

A primal feasible $x$ and a dual feasible $Z$ are optimal, if and only if

$$F(x) \cdot Z = Z \cdot F(x) = 0 \quad (3\text{-}13)$$

This latter condition is called the complementary slackness condition [6]. Substituting (3-8) in slackness condition (3-13) implies that $s$ has to satisfy both of the following homogenous linear equations.

$$(sI + \mathbf{1}\mathbf{1}^T/n - W) \cdot z_1 = 0 \quad (3\text{-}14\text{-a})$$

$$(sI - \mathbf{1}\mathbf{1}^T/n + W) \cdot z_2 = 0 \quad (3\text{-}14\text{-b})$$

Multiplying both sides of equations (3-14) by $\mathbf{1}\mathbf{1}^T$, one can conclude that

$$\mathbf{1}\mathbf{1}^T z_1 = 0 \tag{3-15-a}$$

$$\mathbf{1}\mathbf{1}^T z_2 = 0 \tag{3-15-b}$$

To have the strong duality [6] we set $c^T x + Tr[F_0 Z] = 0$, hence we have

$$s = z_1^T z_1 - z_2^T z_2 \tag{3-16}$$

Considering the linear independence of $\alpha_i, i = 1, \ldots, n-1$, we can expand $z_1$ and $z_2$ in terms of $\alpha_i$ as

$$z_1 = \sum_{i=1}^{n-1} a_i \alpha_i \tag{3-17-a}$$

$$z_2 = \sum_{i=1}^{n-1} a'_i \alpha_i \tag{3-17-b}$$

with the coordinates $a_i$ and $a'_i$, $i = 1, \ldots, n-1$ to be determined. Using (3-3) and the expansions (3-17), while considering (3-15), the slackness conditions (3-14), can be written as

$$(-s + 1) a_i = 2 w_i \alpha_i^T z_1 \tag{3-18-a}$$

$$(s + 1) a'_i = 2 w_i \alpha_i^T z_2 \tag{3-18-b}$$

Considering (3-11) and (3-18), we obtain

$$(-s + 1)^2 a_i^2 = (s + 1)^2 a_i'^2 \tag{3-19}$$

or equivalently

$$\frac{a_i^2}{a_j^2} = \frac{a_i'^2}{a_j'^2} \tag{3-20}$$

For $\alpha_i^T z_1$ and $\alpha_i^T z_2$, we have

$$\alpha_i^T z_1 = \sum_{j=1}^{n-1} a_j G_{i,j} \tag{3-21-a}$$

$$\alpha_i^T z_2 = \sum_{j=1}^{n-1} a'_j G_{i,j} \tag{3-21-b}$$

where $G$ is the Gram matrix, defined as

$$G_{i,j} = \alpha_i^T \alpha_j$$

or equivalently

$$G = \frac{1}{2}\begin{bmatrix} 2 & -1 & 0 & \cdots & 0 \\ -1 & 2 & -1 & \cdots & 0 \\ 0 & -1 & 2 & \ddots & \vdots \\ \vdots & \vdots & \ddots & \ddots & -1 \\ 0 & 0 & \cdots & -1 & 2 \end{bmatrix} \tag{3-22}$$

Substituting (3-21) in (3-18) we have

$$(-s + 1 - 2w_1) a_1 = -w_1 a_2 \tag{3-23-a}$$

$$(-s + 1 - 2w_i) a_i = -w_i (a_{i-1} + a_{i+1}) \tag{3-23-b}$$

$$(-s + 1 - 2w_{n-1}) a_{n-1} = -w_{n-1} a_{n-2} \tag{3-23-c}$$

and

$$(s + 1 - 2w_1) a'_1 = -w_1 a'_2 \tag{3-24-a}$$

$$(s + 1 - 2w_i) a'_i = -w_i (a'_{i-1} + a'_{i+1}) \tag{3-24-b}$$

$$(s + 1 - 2w_{n-1}) a'_{n-1} = -w_{n-1} a'_{n-2} \tag{3-24-c}$$

where (3-23-b) and (3-24-b) are true for $i = 2, \ldots, n-2$.
We are going to determining $s$ (*SLEM*), the optimal weights and the coordinates $a_i$ and $a'_i$, in an inductive manner as follows:
In the first stage, from comparing equations (3-23-a) and (3-24-a) and considering the relation (3-20), we can conclude that

$$(-s + 1 - 2w_1)^2 = (s + 1 - 2w_1)^2 \tag{3-25}$$

which results in $w_1 = 1/2$ and $s = 0$, where the latter is not acceptable.
Assuming $s = \cos(\theta)$ and substituting $w_1 = \frac{1}{2}$ in (3-23-a) and (3-24-a), we may write

$$a_2 = \frac{\sin(2\theta)}{\sin(\theta)} a_1 \tag{3-26-a}$$

$$a'_2 = \frac{\sin(2(\pi - \theta))}{\sin(\pi - \theta)} a'_1 \tag{3-26-b}$$

For the $i$-th stage, assuming $a_j = \frac{\sin(j\theta)}{\sin(\theta)} a_1$ and $a'_j = \frac{\sin(j(\pi - \theta))}{\sin(\pi - \theta)} a'_1$ for $\forall j \leq i$, while considering the relation (3-20), from equations (3-23-b) and (3-24-b) we get

$$\left( (-s + 1 - 2w_i) \frac{\sin(i\theta)}{\sin(\theta)} + w_i \frac{\sin((i-1)\theta)}{\sin(\theta)} \right) a_1 = -w_i a_{i+1} \tag{3-27}$$

$$\left( (s + 1 - 2w_i) \frac{\sin(i(\pi - \theta))}{\sin(\pi - \theta)} + w_i \frac{\sin((i-1)(\pi - \theta))}{\sin(\pi - \theta)} \right) a'_1 \tag{3-28}$$
$$= -w_i a'_{i+1}$$

while considering relation (3-20) we can conclude that

$$\left( (-s + 1 - 2w_i) \sin(i\theta) + w_i \sin((i-1)\theta) \right)^2$$
$$=$$
$$\left( (s + 1 - 2w_i) \sin(i(\pi - \theta)) + w_i \sin((i-1)(\pi - \theta)) \right)^2$$

which results in

$$w_i = \frac{1}{2} \tag{3-29}$$

and $s = 0$, where the latter is not acceptable. Substituting $w_i = 1/2$ in (3-27) and (3-28), we have

$$a_{i+1} = \frac{\sin((i+1)\theta)}{\sin(\theta)} a_1 \qquad (3\text{-}30\text{-a})$$

$$a'_{i+1} = \frac{\sin(2(\pi - \theta))}{\sin(\pi - \theta)} a'_1 \qquad (3\text{-}30\text{-b})$$

where (3-29) and (3-30) are true for $i = 1, \ldots, n-2$ and in the $(n-1)$-th stage, from comparing equations (3-23-c) and (3-24-c) while considering relation (3-20) and using (3-30) for $i = n-2$, we can conclude that

$$w_{n-1} = \frac{1}{2} \qquad (3\text{-}31)$$

Then by substituting $w_{n-1} = 1/2$ in one of equations (3-23-c) and (3-24-c), we obtain
$$\sin(n\theta) = 0$$

which in turn results in $\theta_k = k\pi/n$ for $k = 1, \ldots, n-1$, therefore *SLEM* equals largest root of $s$ in magnitude which is $cos(\pi/n)$, the results thus obtained are in agreement with those of [3]. Also one should notice that necessary and sufficient conditions for the convergence of weight matrix are satisfied, since all roots of $s$ which are the eigenvalues of $W$ are strictly less than one in magnitude, and one is a simple eigenvalue of $W$ associated with the eigenvector **1** where this happens due to positivity of optimal weights [1].

Finally dual constraints (3-11), (3-12) and strong duality condition (3-16) are satisfied for $s = cos(\pi/n)$, the optimal weights (3-29) and (3-31), the coordinates (3-30) and

$$a_1 = \frac{1 + \cos(\theta)}{1 - \cos(\theta)} \times \frac{1}{n} \times \sin^2\theta \qquad (3\text{-}32\text{-a})$$

$$a_1 = \frac{1 - \cos(\theta)}{1 + \cos(\theta)} \times \frac{1}{n} \times \sin^2\theta \qquad (3\text{-}32\text{-b})$$

## IV. Conclusion

Here in this paper we have proposed a practical and versatile method for analytical solution of finding optimal weights that yield the fastest convergence rate to the asymptotic solution by means of SDP, for the Path network. Our approach is based on fulfilling the slackness conditions where the optimal weights are obtained by inductive comparing of the resultant characteristic polynomials initiated via slackness conditions. We believe that this method is powerful and lucid enough to be extended to other networks with more general topologies.